\documentclass[twocol]{ametsocV5}

\usepackage{amsmath,amsfonts,amssymb,bm}
\usepackage{mathptmx}
\usepackage{newtxtext}
\usepackage{newtxmath}
\usepackage[version=4]{mhchem}
\usepackage[inline]{enumitem}
\usepackage{csquotes}
\usepackage{siunitx}
\usepackage{dutchcal}

\newcommand{\backvec}[1]{\reflectbox{$\vec{\reflectbox{\!$#1$}}$}}

\newcommand{\ddx}[2]{\frac{\text{d} #1}{\text{d} #2}}
\newcommand{\drdrx}[2]{\frac{\partial #1}{\partial #2}}

\newcommand{\tc}{\mathscr{t}_{\mathrm{c}}}
\newcommand{\lc}{\mathscr{l}_{\mathrm{c}}}
\newcommand{\tev}{\mathscr{t}_{\mathrm{e}}}
\newcommand{\lev}{\mathscr{l}_{\mathrm{e}}}
\newcommand{\lD}{\mathscr{l}_\Delta}
\newcommand{\xD}{\overline{x}^\Delta}
\newcommand{\yD}{\overline{y}^\Delta}

\title{A minimal model of the deep-convection lifecycle and its verification in remote-sensing observations}

\authors{Tobias Bölle\correspondingauthor{Tobias Bölle, tobias.boelle@dlr.de}$^\dagger$,
        Christoph Metzl\thanks{These authors contributed equally.} and 
        Kianusch Vahid Yousefnia}

\affiliation{Deutsches Zentrum für Luft- und Raumfahrt, Institut für Physik der Atmosphäre, Oberpfaffenhofen, Germany}

\abstract{Deep convection is one of the most important atmospheric transport mechanisms and associated with various severe weather phenomena. Manifestations of deep convection in the atmosphere are composed of a recurring fundamental building block, called cell, which evolves through a characteristic lifecycle. Despite its importance, no simple, physically consistent quantitative lifecycle model exists that correctly reproduces remote-sensing observations. Based on the standard conceptual model, we develop an analytic minimal model of the convection lifecycle in the form of coupled reaction rules. This reaction scheme is equivalent to a system of coupled nonlinear differential equations that qualitatively agree with the empirically known dynamics. In order to demonstrate quantitative agreement of our convection-lifecycle model, we construct a representative random sample of satellite and radar observations of deep-convection events over Germany. In particular, we introduce a conditional sampling strategy based on an objective timescale criterion to remove events not covered by our model from the random sample. We show that by this procedure, a definite and recurrent temporal signature of the convection lifecycle can be identified. Our model closely reproduces the principal characteristics while essentially staying within one standard deviation of the mean signature. Next to the definite relevance of our model for nowcasting, the approach may be useful for convection parametrisations.}

\begin{document}
\maketitle

%
%
\statement
In this study, we develop a quantitative minimal model for deep convection, manifesting as thunderstorms, which are associated to severe weather. To mitigate the societal and economic impact, highly accurate and timely forecasts are critical. This makes a reliable, quantitative minimal model of the governing deep-convection dynamics necessary. In this study, we develop such a minimal model by formalising empirical prior knowledge. We then show that our model correctly captures the essential evolution of thunderstorms observed in satellite imagery and radar data, on average. The model extends recent advancements in short-term forecasting and convection parametrisation, suggesting potential benefits for forecasting on various lead times. Our work contributes to the broader understanding and prediction of thunderstorms.

%

%




\section{Introduction} 
\label{sec:Introduction}

Deep convection is one of the most important and efficient vertical transport mechanisms of momentum, energy and moisture across the troposphere \citep{Lilly1979,Doswell1985,Stevens2005}. Thus, next to being a canonical atmospheric process of fundamental physical interest \citep{Emanuel1994}, it crucially impacts the earth energy budget through transport and radiation \citep{Sherwood2010}. Furthermore, deep convection, roughly synonymous with a thunderstorm, is associated with several hazardous weather phenomena, such as heavy rain, hail, strong winds and lightning \citep{Doswell2001,Markowski2010}. This implies a direct societal and economic impact, making highly accurate and timely short-term forecasts necessary \citep{Wang2017,Bojinski2023,Zhang2023,Vahid2024}. Improving our understanding and efficiently modelling deep convection are therefore crucial to improve model parametrisations, 
climate projections \citep{Arakawa2004,Wagner2010,Sullivan2023} 
and short-term forecasting of hazardous weather \citep{Sun2014,Wapler2021,Wilhelm2023,Vahid2024b}. \par 

While deep convection tends to organise spatiotemporally in recurrent patterns, there is definite observational evidence that these patterns are essentially composed of a universal building block, called (deep-convection or thunderstorm) cell \citep{Lilly1979,Doswell1985,Houze1993}. 
The objective of this study is to develop a minimal model that captures the essential features and evolution of a prototypical, isolated deep-convection cell. The integral dynamics of a deep-convection cell is referred to as its lifecycle \citep{Doswell1985,Markowski2010}. \par
While the cell lifecycle is \emph{qualitatively} well understood, there exists no simple, physically consistent model for its \emph{quantitative} description to the best of our knowledge \citep{Doswell1985,Emanuel1994,Markowski2010,Yano2012,Colin2021,Wilhelm2023}. Instead, modelling approaches so far concentrated on various aspects of the whole lifecycle. This way, considerable understanding was gained from studying parcel, plume and bubble models \citep{Houze1993,Markowski2010,Sherwood2013,Yano2014,Romps2015,Hernandez-Deckers2016}, density-currents \citep{Emanuel1994}, coupled cloud-layer models \citep{Arakawa2004,Wacker2006,Davies2009} as well as particle-based lattice models \citep{Bengtsson2013,Boing2016}. Especially in nowcasting, lifecycle models are typically inferred from statistical analyses and are thus not necessarily physically consistent \citep{Weusthoff2008,Pulkkinen2019,Wapler2021,Wilhelm2023,Zhang2023}. \par

Recently, population-dynamics approaches similar to the Lotka--Volterra system have proven successful for the qualitative understanding of cloud dynamics \citep{Koren2011,Gjini2023,Gordon2023,Glassmeier2024}. These models assume the existence of small sets of variables that characterise the essential processes governing the dynamics of individual clouds or cloud systems \citep{Feingold2013}. We refer to these variables as \emph{macroscopic} to distinguish them from the \emph{microscopic} description necessary to resolve all ongoing cloud processes \citep{Pujol2019,Colin2021}. These studies showed that the macroscopic dynamics of convective clouds is well described by a set of coupled nonlinear oscillators \citep{Wacker2006,Feingold2013,Koren2017,Pujol2019}. Previous work mostly concerned regime switching between open- and closed-cellular patterns of marine stratocumulus clouds \citep{Koren2011,Pujol2019,Gordon2023}. However, similar approaches have been applied successfully to qualitatively model aspects of deep-convective processes \citep{Colin2021} and proposed to improve convection parametrisation \citep{Wagner2010,Yano2012}. \par

Population dynamics is essentially equivalent to reactions between different species \citep{Falk1968,Haken1978}. A systematic way to develop dynamical systems of interacting (or reacting) species is by formulating appropriate reaction rules \citep{Prigogine1967,Gillespie2007}. This approach has a certain similarity with rule-based expert systems to infer thunderstorm occurrence in multi-data ensembles used in short-term forecasting \citep{Wang2017}. Conceptually, reaction rules may be understood to extend and quantify the ingredients-based method of \citet{Doswell1996} which has great importance in operational forecasting. However, to the best of our knowledge, this approach has not been used to develop deep-convection models before. Thus, the first objective of this study is to show the usefulness of this framework for a systematic model development on the basis of empirical prior knowledge. By construction, the resulting model \emph{qualitatively} agrees with the known behaviour used to build it. \par 

In order for these macroscopic cloud-dynamics models to have a true added value, they must also \emph{quantitatively} characterise actual realisations of the phenomenon. Previous studies used satellite imagery essentially to motivate the approach, focusing on qualitative agreement between the model behaviour and observational data \citep{Koren2011,Gordon2023}. Also, model comparison against large-eddy simulations was essentially on a qualitative level \citep{Colin2021}. We are not aware of any attempt at quantitatively relating population dynamics-like models to remote-sensing observations of deep-convection cloud dynamics. Hence, the second objective of this study is to demonstrate the capacity of our macroscopic cloud model to qualitatively and quantitatively match remote-sensing observations of atmospheric deep convection. \par

The main result of this study is a minimal model for the macroscopic description of the convection lifecycle. Working in the framework of reaction schemes guarantees consistency with empirical prior knowledge. We further propose a strategy to define and sample corresponding lifeycle time series from satellite imagery and radar data. Our model correctly reproduces the governing features of these observational signatures. Eventually, we show agreement with a model recently inferred from statistical radar data analysis to describe the mature stage of the lifecycle \citep{Wapler2021,Wilhelm2023}. Our model extends the temporal range of validity to earlier and later development stages and the use of satellite imagery. \par

The present study provides a proof of concept that a quantitatively accurate convection-lifecycle model can be developed in the framework of reaction schemes. Nevertheless, our results have direct relevance for short-term forecasting of deep convection, particularly in the scope of hybrid physics-informed data-driven model building \citep{Pulkkinen2019,Reichstein2019}. Reaction schemes provide a very flexible framework for systematic model development from empirical knowledge. Hence, in principle, our model may be readily extended to show other process behaviour or couple to further meteorological variables, which may be necessary to account for convection organisation. Our first results together with this flexibility suggest usefulness in the context of model parametrisations that lack important process details of deep convection \citep{Arakawa2004,Davies2009,Wagner2010}. \par

The paper is organised as follows. Motivated by conceptual models and observational evidence, we develop and discuss a quantitative model of the convection lifecycle in section~\ref{sec:Model}. We then introduce the relevant observational data and post processing to evaluate the model quality in section~\ref{sec:Data}. Eventually, section~\ref{sec:Model_observation_comparison} contains a quantitative comparison between the model and remote-sensing observations. We conclude our results in section~\ref{sec:Conclusion}.

\section{Model} 
\label{sec:Model}

Deep convection is the consequence of the coherent interaction of a multitude of different dynamical, thermodynamical and microphysical processes \citep{Stevens2005,Feingold2013}. The details of these processes are intricate and not fully understood yet, so that a detailed \textquote{microscopic} description is impossible. However, it is well known that for systems of many interacting elements a \textquote{macroscopic} description of the essential dynamics is often possible \citep{Haken1978,Pujol2019}. In this work, we assume that the governing deep-convection dynamics can be described in terms of particle-like variables. Interactions in this case are formally equivalent to reactions between chemical species \citep{Falk1968}.

\subsection{Model development}
\label{sec:Reaction_model}
Convection is the consequence of a released instability and tends to redistribute the atmospheric fluid in a way to reestablish a stable state \citep{Emanuel1994,Doswell2001}. 
This net effect of convection to modify atmospheric fields in a stabilising manner forms the conceptual basis of a family of convection parametrisation \citep{Arakawa2004}. 
We formalise this fact by introducing the overall reaction 
\begin{equation}\label{eq:Global_reaction}
    \ce{$A$ <=>[$\vec{\varphi}$][$\backvec{\varphi}$] $Z$} 
\end{equation}
associated with convection. Herein, $A$ and $Z$ denote reservoirs containing the (species of) \textquote{active} atmospheric state, in which convection can principally develop, and the \textquote{inactive} state of the atmosphere exhausted by convection, respectively. In meteorological terms, $A$ corresponds to the (large-scale) environmental processes that exchange directly with deep convection.
Let $\lc$ ($\tc$) and $\lev$ ($\tev$) denote the characteristic length (time) scales of the deep-convection lifecycle and large-scale environment, respectively. Considering the environment as a reservoir for the deep-convection processes implies that $\lc \ll \lev$ and $\tc \ll \tev$. The reaction~\eqref{eq:Global_reaction} principally runs in both directions and $\vec{\varphi}, \backvec{\varphi} \ge 0$ denote the corresponding kinetic constants. \citet{Yano2012} refer to this as discharge--recharge mechanism. \par 
There are two ways for the reaction \eqref{eq:Global_reaction} to be stationary. First, thermodynamic equilibrium corresponds to the distinguished state (extremum of the relevant thermodynamic potential) in which the forward and backward running reactions precisely balance \citep{Sommerfeld1952,Falk1968}. Second, it is principally conceivable to supply $A$ and remove $Z$ at a rate to establish a stationary non-equilibrium state \citep{Haken1978}. In fluid mechanics, these would be the laminar and turbulent flow regimes. For a stationary turbulent state production equals dissipation \citep{Tennekes1972}. Such a state is approximately realised for synoptic scales \citep{Monin1972} and constitutes a useful concept in the statistical theory of turbulence (and convection) \citep{Emanuel1994b}. However, it may not hold for individual processes or events. Emphasising the importance of such individual processes and their interaction is the coherent-structures paradigm \citep{Holmes1996}. \par

In the scope of our model, the actually realised atmospheric processes correspond to the sense in which reaction \eqref{eq:Global_reaction} is running. Using standard notation, the forward and backward reaction rates, $\vec{v}$ and $\backvec{v}$, are symbolised by arrows indicating the reaction sense. (We emphasise that $\vec{v}$ and $\backvec{v}$ are scalar-valued quantities.) Assuming that reaction rates are proportional to the species amount, we can express $\vec{v} = \vec{\varphi} a$ and $\backvec{v} = \backvec{\varphi} z$, where $a, z$ denote the concentrations associated with $A, Z$ \citep{Haken1978}. The kinetic expression of the global reaction rate then reads
\begin{equation}\label{eq:Global_reaction_rate_1}
    v = \vec{v} - \backvec{v} = \vec{v}\left(1 - \frac{\backvec{\varphi}}{\vec{\varphi}}\frac{z}{a}\right) .
\end{equation}

The kinetic constants $\vec{\varphi}$ and $\backvec{\varphi}$ appearing in \eqref{eq:Global_reaction_rate_1} can be related to the fundamental concept of affinity, measuring how far a reaction is from equilibrium \citep{Prigogine1967}. The affinity of the global reaction \eqref{eq:Global_reaction} reads 
\begin{equation}\label{eq:Global_affinity}
    \mathcal{A} = RT \ln \frac{K(T)}{a^{-1}z} ,
\end{equation}
assuming that the system is in contact with a heat bath of energy $RT$. By definition, $\mathcal{A} = 0$ at equilibrium, which, by \eqref{eq:Global_affinity} implies that the ratio of active to inactive environmental state $a^{-1}z$ must equal the equilibrium constant $K$; this is the the law of mass action \citep{Sommerfeld1952,Falk1968}. Non-equilibrium is defined as $\mathcal{A} > 0$ and depends on the relative concentrations of the environment. Since the equilibrium constant $K$ equals the ratio of kinetic constants $\vec{\varphi}\,\backvec{\varphi}^{-1}$  \citep{Prigogine1967}, inserting \eqref{eq:Global_affinity} into \eqref{eq:Global_reaction_rate_1} yields
\begin{equation}\label{eq:Global_reaction_rate_2}
    v= \vec{v}\left(1 - \exp{\left(-\frac{\mathcal{A}}{RT}\right)}\right) .
\end{equation}
Thinking of the affinity as the energy required to start the reaction \citep{Sommerfeld1952,Falk1968,Prigogine1967} and the heat bath as a measure of the fluctuations, \eqref{eq:Global_reaction_rate_2} expresses the probability that the reaction \eqref{eq:Global_reaction} runs in either possible direction. This has striking similarities with the concept of convective inhibition \citep{Markowski2010}. \par
Atmospheric states close to equilibrium correspond to $|\mathcal{A}(RT)^{-1}| \ll 1$, which, by Taylor-series expansion of \eqref{eq:Global_reaction_rate_2}, yields the global reaction rate $(RT)^{-1}\mathcal{A}\vec{v}_\mathrm{eq}$, if $\vec{v}_\mathrm{eq}$ denotes the partial reaction rate at equilibrium. In this case, the atmosphere is in a state of quasi-equilibrium in which the destabilisation of the large-scale environment is approximately balanced by convective stabilisation and $\tc \sim \tev$. This is characteristic of the tropics \citep{Doswell2001}. \par 
On the other hand, the atmospheric state far from equilibrium is associated with $\mathcal{A}(RT)^{-1} \to \infty$, which implies, by \eqref{eq:Global_reaction_rate_2}, that $v \to \vec{v}$. 
This implies a timescale separation $\tc \ll \tev$ between the environmental large scales and convection. Hence, on the scale of the large scales, convection is always close to stationary. This convective quasi-equilibrium hypothesis forms the basis of convection parametrisations by adiabatic elimination \citep{Haken1978,Emanuel1994,Yano2012c}. A state, in which the large-scale environment is far from equilibrium, is typically found in the midlatitudes \citep{Doswell2001}. \par 

In what follows, we assume timescale separation $\tc \ll \tev$ to hold so that the reaction~\eqref{eq:Global_reaction} runs only to the right. This is formally equivalent to considering only the transient initial phase of the whole reaction. \par

\begin{figure*}
    \centering
    \includegraphics[width=.9\textwidth]{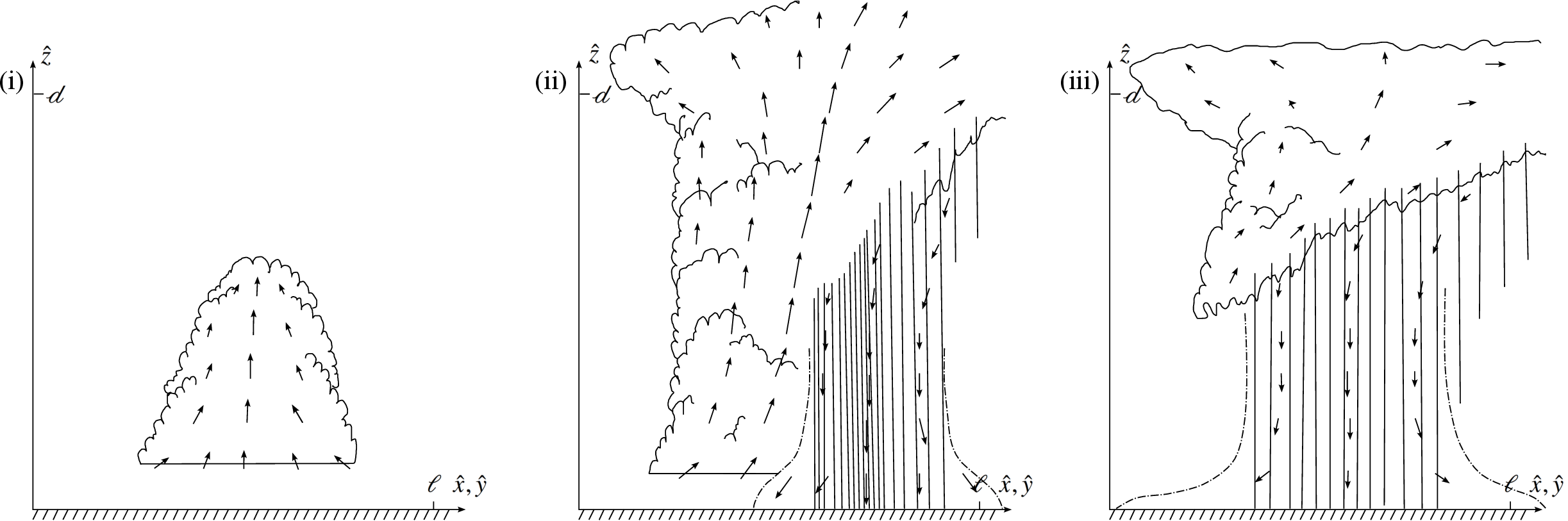}
    \caption{Schematic of the conceptual deep-convection lifecycle (reproduced after \citealt{Doswell1985}). Deep convection, manifesting as cumulonimbus clouds in the atmosphere, consecutively pass through their three development stages: (i) towering cumulus, (ii) mature and (iii) dissipating. The characteristic length scales are $\mathscr{d} \sim \mathscr{l} \sim \SI{10}{\km}$ and the Cartesian coordinate system is denoted $(\hat{x}, \hat{y}, \hat{z})$.}
    \label{fig:Lifecycle}
\end{figure*}
We now argue that this part of the global reaction~\eqref{eq:Global_reaction} actually consists of intermediate reactions corresponding to the deep-convection lifecycle. The concept of cell lifecycle is closely related to the manifestation of deep convection as cumulonimbus clouds \citep{Doswell1985,Houze1993,Sullivan2023}. According to common understanding, a prototypical isolated deep-convection cell consecutively passes three stages: the 
\begin{enumerate*}[label=(\roman*)]
    \item towering-cumulus,
    \item mature, and
    \item dissipating stage, 
\end{enumerate*}
schematically reproduced from \citet{Doswell1985} in figure~\ref{fig:Lifecycle}. \par
The distinctive characteristics of cumulonimbus clouds are the vertically heaping cloud structure and resulting strong precipitation
\citep{Doswell1985,Houze1993}. The distinctive role of precipitation in characterising deep convection was already stressed by \citet{Hernandez-Duenas2013}. This is also reflected in figure~\ref{fig:Lifecycle}, as the entire lifecycle can be schematically composed of these two elements. Consequently, next to $A$ and $Z$, we introduce $X$ and $Y$ to describe the \textquote{cumulus-convection} and \textquote{precipitation} species, respectively. A similar choice of variables has proven successful in modelling strato-cumulus dynamics \citep{Koren2011,Pujol2019}. These variables exchange directly with the active environment $A$. We then propose the following detailed reactions 
\begin{align}
    \ce{$A$ + $X$ ->[$\alpha$] $2X$} \label{eq:X_2X} \\
    \ce{$X$ + $Y$ ->[$\beta$] $2Y$} \label{eq:X+Y_2Y} \\
    \ce{$X$ + $Z$ ->[$\gamma$] $2Z$} \label{eq:X+Z_2Z} \\
    \ce{$Y$ + $Z$ ->[$\delta$] $2Z$} \label{eq:Y+Z_2Z}
\end{align}
to model the deep-convection lifecycle. Reaction~\eqref{eq:X_2X} implies that existing cumuli tend to become deeper in an autocatalytic, self-amplifying manner as long as energy is available. Precipitation forms if cumulus clouds grow sufficiently deep. Given the existence of an (infinitesimal) amount of precipitation, it will amplify in the autocatalytic reaction~\eqref{eq:X+Y_2Y} as long as the feeding cumulus cloud persists. In the context of population dynamics, this feedback is referred to as a predator--prey relationship \citep{Koren2011}. Eventually, both the growing cumulus and precipitation species tend to produce inactive state in reactions~\eqref{eq:X+Z_2Z} and \eqref{eq:Y+Z_2Z}. This reflects the stabilising effect of both convective up- and downdraughts \citep{Doswell2001}. In the following, we assume that $A$ is held constant. Deep convection then corresponds to an open system that is maintained in a state far from equilibrium: the constantly supplied reactant $A$ is transformed in the reaction \eqref{eq:X_2X}--\eqref{eq:Y+Z_2Z} to the residual $Z$. \par
Analogous to \eqref{eq:Global_affinity}, we can introduce the affinity 
$\mathcal{A}_r := -\sum_j \nu_{rj} \mu_j$ 
of the $r$th reaction in \eqref{eq:X_2X}--\eqref{eq:Y+Z_2Z}. Since all intermediate states are unstable, the affinity of the global reaction becomes $\mathcal{A} = \sum_r \mathcal{A}_r$ \citep{Prigogine1967}. \par 
As deep convection is associated with turbulent dynamics, we can expect the various constituting elements to be thoroughly mixed \citep{Emanuel1994,Yano2014}. Consequently, spatial variability smaller than the cloud scale is always rapidly and efficiently smoothed out. We can therefore consider the system to be uniformly randomised in space. For this well-mixed system, the reactions~\eqref{eq:X_2X}--\eqref{eq:Y+Z_2Z} are equivalent to the coupled system of nonlinear ordinary differential equations \citep{Gillespie2007} 
\begin{align}
    \ddx{x}{t} &= \alpha x - \beta xy - \gamma zx \label{eq:dxdt} \\
    \ddx{y}{t} &= \beta xy - \delta yz \label{eq:dydt} \\
    \ddx{z}{t} &= \gamma zx + \delta zy \label{eq:dzdt} , 
\end{align}
where the respective lower-case letters denote concentrations corresponding to the species. (The environmental species $A$ is absorbed in the kinetic coefficient $A\alpha \to \alpha$.) Subject to nonsingular initial conditions (i.e. no fixed point), a locally unique solution to \eqref{eq:dxdt}--\eqref{eq:dzdt} exists \citep{Arnold1985}. 
Comparing the reactions~\eqref{eq:X_2X}--\eqref{eq:Y+Z_2Z} with the associated differential equations~\eqref{eq:dxdt}--\eqref{eq:dzdt} shows that two-species reactions correspond to quadratic nonlinearities. The nonlinearities imply that the cloud has to reach considerable (finite) depth to produce significant precipitation. Also, only deep and strongly precipitating clouds lead to significant production of inactive species and, hence, stabilisation. \par

\subsection{Model discussion}
\label{sec:Model_discussion}

\begin{figure}
    \centering
    \noindent\includegraphics[width=.9\columnwidth]{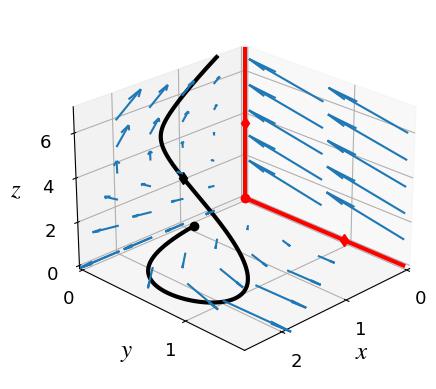}   
    \caption{Model dynamics in $(x,y,z)$ phase space ($\alpha = 0.29$, $\beta = 0.30$, $\gamma = 0.08$, $\delta = 0.52$). Straight lines and dot show the model fixed points with diamond symbols indicating the critical values of changing stability regimes. Arrows display projections of the vector field $\bm{f}(\bm{x})$ which coincide with the eigenspaces in the vicinity of the fixed points. The parametric curve shows an exemplary model trajectory started from $\bm{x}_0 = (1,0.1,0.1)$. The initial condition is indicated by a dot and the diamond marker indicates the critical time $t_\mathrm{c} = \alpha\gamma^{-1}$.}
    \label{fig:Phase_space}
\end{figure}

Writing $\bm{x} = (x,y,z) = (x_1,x_2,x_3)$, such that $x_i \in \mathbb{R}^+$ for $i = 1,2,3$, the model system \eqref{eq:dxdt}--\eqref{eq:dzdt} has the symbolic form 
\begin{equation}\label{eq:Formal_ode}
    \ddx{\bm{x}}{t} = \bm{f}(\bm{x}) ,\quad \bm{x}(t=0) = \bm{x}_0 ,
\end{equation}
of a vector-valued, autonomous differential equation. The right-hand side of \eqref{eq:Formal_ode} is a vector field parameterised by $(\alpha, \beta, \gamma, \delta) > 0$. For the specific parameter setting $\alpha = 0.29$, $\beta = 0.30$, $\gamma = 0.08$, $\delta = 0.52$ (section~\ref{sec:Model_observation_comparison}), figure~\ref{fig:Phase_space} shows the three-dimensional trajectory in phase space starting from the initial condition $\bm{x}_0 = (1.0,0.1,0.1)$. \par 

The fixed points of \eqref{eq:Formal_ode}, i.e. solutions $\bm{x}^*$ such that $\bm{f}(\bm{x}^*) = 0$, are
\begin{equation}\label{eq:Fixed_points}
    \bm{x}^*_1 = 
    \begin{pmatrix}
        0 \\ 0 \\ 0
    \end{pmatrix} ,\quad
    \bm{x}^*_2 = 
    \begin{pmatrix}
        0 \\ y \\ 0
    \end{pmatrix} ,\quad
    \bm{x}^*_3 = 
    \begin{pmatrix}
        0 \\ 0 \\ z
    \end{pmatrix} .
\end{equation}
The first and last fixed points $\bm{x}^*_1$ and $\bm{x}^*_3$ directly express expected dynamical behaviour. Namely, that convection cannot proceed from a perfectly perturbation-free atmospheric state and that a completely inactive state will remain so, respectively. The second fixed point $\bm{x}^*_2$, representing perpetual precipitation in the absence of any other species, does not correspond to an atmospheric state which would actually be observed. The fixed points are shown in figure~\ref{fig:Phase_space} in red. \par
Considering the stability of \eqref{eq:Fixed_points} in the following, we show that $\bm{x}_1^*$ and $\bm{x}_2^*$ are unstable for physically consistent parameter combinations and, therefore, cannot actually be realised. However, $\bm{x}_3^*$ becomes stable beyond a critical value and constitutes the inevitable endpoint of any deep-convection process.
Asymptotic stability of the fixed points \eqref{eq:Fixed_points} is established by considering the eigenvalues of the linearised vector field \citep{Guckenheimer1983}
\begin{equation}\label{eq:Jacobian}
    \drdrx{f_i}{x_j}\bigg|_{\bm{x}^*} = 
    \begin{bmatrix}
        \alpha - \beta y^* - \gamma z^* & -\beta x^* & -\gamma x^* \\
        \beta y^* & \beta x^* - \delta z^* & -\delta y^* \\
        \gamma z^* & \delta z^* & \gamma x^* + \delta y^*
    \end{bmatrix} .
\end{equation}
Three projections of the vector field on $x_i = 0$ are included in figure~\ref{fig:Phase_space} as blue arrows. For the eigenvalues and -vectors of \eqref{eq:Jacobian}, we obtain
\begin{align}
    \bm{\lambda}(\bm{x}^*_1) &= 
    \begin{pmatrix}
        0 \\
        0 \\
        \alpha \\
    \end{pmatrix} 
    ,\quad &&\bm{E}_1 = 
    \begin{bmatrix}
        0 & 0 & 1 \\ 1 & 0 & 0 \\ 0 & 1 & 0
    \end{bmatrix} , \label{eq:Eigenvals+vecs_FP1} \\
    \bm{\lambda}(\bm{x}^*_2) &= 
    \begin{pmatrix}
        0 \\
        \delta y^* \\ 
        \alpha - \beta y^*
    \end{pmatrix}
    ,\quad &&\bm{E}_2 = 
    \begin{bmatrix}
        0 & 0 & \frac{\alpha-\beta y^*}{\beta y^*} \\ 1 & -1 & 1 \\ 0 & 1 & 0
    \end{bmatrix} , \label{eq:Eigenvals+vecs_FP2} \\
    \bm{\lambda}(\bm{x}^*_3) &= 
    \begin{pmatrix}
        0 \\
        -\delta z^* \\ 
        \alpha - \gamma z^*
    \end{pmatrix}
    ,\quad &&\bm{E}_3 = 
    \begin{bmatrix}
        0 & 0 & \frac{\alpha-\gamma z^*}{\gamma z^*} \\ 0 & -1 & 0 \\ 1 & 1 & 1
    \end{bmatrix} . \label{eq:Eigenvals+vecs_FP3}
\end{align}
The columns of the similarity matrices $\bm{E}_i$ ($i = 1,2,3$) contain the eigenvectors associated with the eigenvalues assembled top-down in the vectors $\bm{\lambda}(\bm{x}_i^*)$. \par
From \eqref{eq:Eigenvals+vecs_FP1}, we see that \eqref{eq:Jacobian} evaluated at $\bm{x}^*_1$ (no convection species present at all) has a pure point spectrum. The first fixed point is neutrally stable to perturbations lying in the span of the first and second eigenvector.  (The degenerate null-eigenvalue has algebraic and geometric multiplicity two.) That is, no linear combination of precipitation and inactive state would produce convection. However, $\bm{x}^*_1$ is unstable (recall, $\alpha > 0$) with respect to (infinitesimal) perturbations of cumulus convection $x$. This instability is apparent in figure~\ref{fig:Phase_space}. \par
Inspecting \eqref{eq:Eigenvals+vecs_FP2} shows that \eqref{eq:Jacobian} evaluated at $\bm{x}^*_2$ (perpetual precipitation) has a point and a continuous spectrum. The second fixed point is neutrally stable to perturbations that merely change the amount of precipitation. Given that $\delta > 0$, it is unstable to perturbations that reciprocally change the precipitation and dissipation species by the same amount. Stability with respect to perturbations modifying the precipitation- and cumulus-species amount depends on the actual values of $\alpha, \beta, y^*$. The dynamics is neutrally stable at the critical value $y^*_\mathrm{c} = \alpha\beta^{-1}$ (figure~\ref{fig:Phase_space}), for which the associated eigenvector degenerates to a perturbation assuming only precipitation species. By \eqref{eq:dxdt}--\eqref{eq:dydt}, the critical value corresponds to the singular parameter setting in which the rate of cumulus creation is exactly balanced by the rate of conversion to precipitation. The critical value separates the unstable $y^* < y^*_\mathrm{c}$ from the stable $y^* > y^*_\mathrm{c}$ regime. Since $\beta y^* > 0$, the associated eigenvectors of the unstable (stable) regime correspond to an increase (decrease) of the cumulus relative to the precipitation species. This change in stability regime accompanied by the changing eigenvector orientation is shown in figure~\ref{fig:Phase_space}. \par
The stable regime requires that the rate of cumulus creation $\alpha$ be inferior to its conversion into precipitation $\beta y^*$. This situation is physically inconsistent, as precipitation can be produced only at the rate at which cumuli increase. We conclude that the fixed-point state $\bm{x}^*_2$ of pure precipitation is unstable to infinitesimal perturbations in the cumulus and dissipation species. This state can, therefore, not be realised in practice and is hence unobservable. \par
From \eqref{eq:Eigenvals+vecs_FP3}, we see that the spectrum of \eqref{eq:Jacobian} evaluated at $\bm{x}^*_3$ (exclusively dissipation species present) has a structure analogous to \eqref{eq:Eigenvals+vecs_FP2}. Neutral stability holds for perturbations that merely perturb the amount of dissipation species. The fixed point is stable with respect to (infinitesimal) perturbations of the precipitation and dissipation species in equal but opposite proportions. As before, the last fixed point admits different stability regimes when perturbed by cumulus species. The unstable $z^* < z^*_\mathrm{c}$ and stable $z^* > z^*_\mathrm{c}$ regimes are separated by the critical value $z^*_\mathrm{c} = \alpha\gamma^{-1}$ for which the dynamics is neutrally stable (figure~\ref{fig:Phase_space}). Instability of the pure dissipation state would require that the rate of cumulus production $\alpha$ exceeds its conversion into dissipation species $\gamma z^*$. Since $\alpha, \gamma > 0$ are fixed values and $z(t)$ is a monotonic, non-negative function of time (cf. \eqref{eq:dzdt}), this implies the existence of a critical time $t_\mathrm{c}$ such that $z(t_\mathrm{c}) = z_\mathrm{c}^*$, beyond which convection inevitably ceases. \par 
This stable fixed point of a completely inactive atmospheric state from which no new convection can develop is the endpoint of the convection lifecycle we expect to find in observations. On the other hand, the fact that the fixed point principally admits two distinct stability regimes could be a starting point to introduce mesoscale convective organisation into the model. This would require model adjustments to obtain a more complex metastable fixed-point structure along with additional processes that account for the regime-switching dynamics. \par
We note that the functional characteristics of $z$ being a monotonic, non-negative function has striking resemblance with an entropy. This corroborates our interpretation of $z$ to characterise dissipation. Moreover, this interpretation is consistent with the structure of \eqref{eq:dzdt}, in which changes of $z$ are due to bilinear forms. This is the typical form of the entropy equation in non-equilibrium thermodynamics \citep{Prigogine1967}. As such, we conjecture that $z$ constitutes a measure of irreversibility associated with deep convection and serves to measure progress of the process.







\section{Data} 
\label{sec:Data}

For the model development in section~\ref{sec:Model}, we assumed that deep convection constitutes a coherent flow entity in the sense that a small set of macroscopic variables can be found for its characterisation. On physical reasoning, we then proposed a system of nonlinearly coupled differential equations to determine the temporal evolution of these macroscopic variables. We maintain that this system is a model of the prototypical convection lifecycle as it occurs in real atmospheric flow. Deep convection manifests as cumulonimbus clouds in the atmosphere \citep{Doswell1985,Houze1993} and is therefore principally observable. We conjecture that (the manifestation of) deep convection is associated with a recurrent signature in the relevant observational data and that this signature obeys our model (section~\ref{sec:Model}). 
In this section, we discuss the relevant remote-sensing data and the necessary post-processing of these data to test this hypothesis. \par

\begin{figure}
    \centering
    \includegraphics[width=.48\textwidth]{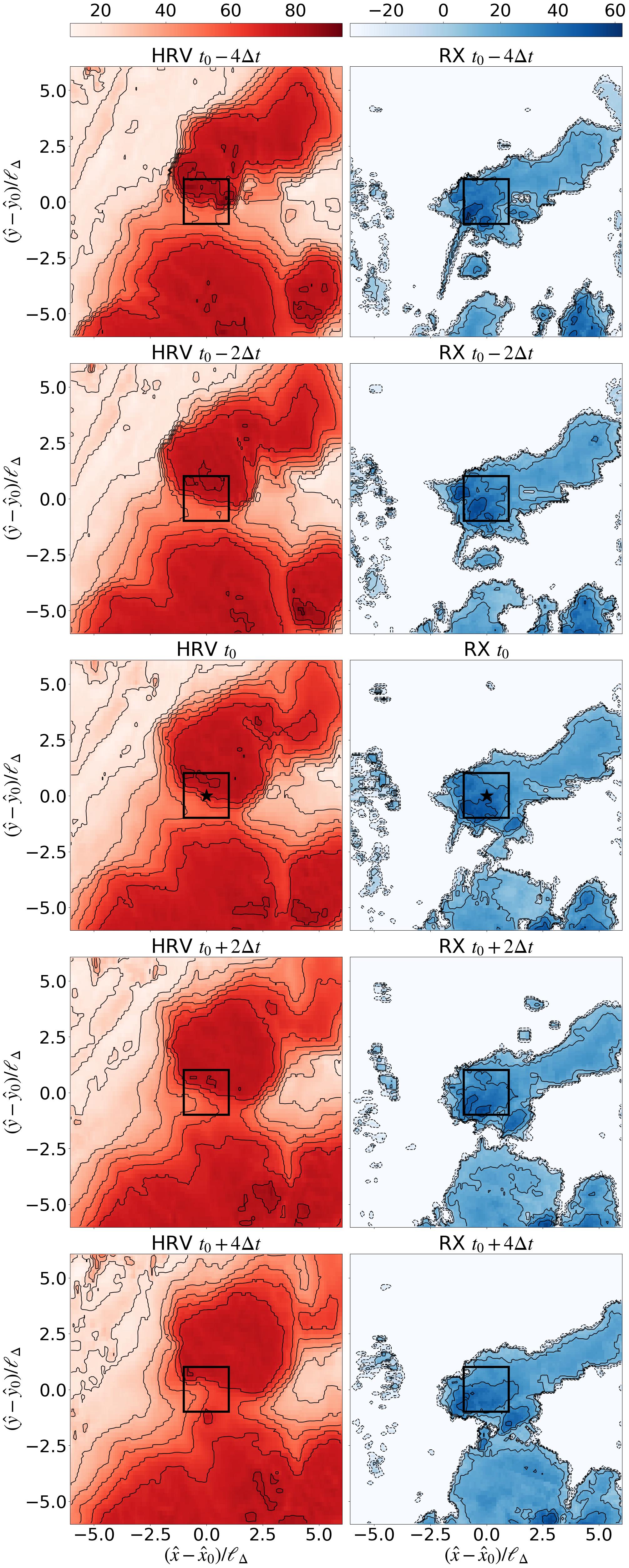}
    \caption{Example convection lifecycle observation. Left column: high resolution visible (HRV) imagery in units of $\si{\percent}$ from the Meteosat Second Generation satellite. Right column: radar reflectivity in units of dBZ from the RX composite over Germany. The thunderstorm event-defining lightning instance (indicated by an asterisk) marks $(t_0,\hat{\bm{x}}_0)$ and subsequent snapshots are separated by $2 \Delta t \tc^{-1} = 0.33$. The box with linear dimension $\mathscr{l}_\Delta\lc^{-1} = 2$ defines the area taken for spatial averaging.}
    \label{fig:RSexample}
\end{figure}

\subsection{Procedure to obtain time series from field data}
\label{sec:Time_series_construction}
On physical reasoning, in section~\ref{sec:Model}, we tentatively identified the macroscopic model variables $x, y$ with cumulus convection and precipitation, respectively. In this study, we use satellite imagery and radar data to associate these (abstract) model variables with measurable quantities. In particular, the observational associates for $x$ and $y$ will be derived from satellite imagery and radar data, respectively. As the model variables are thought to characterise deep convection macroscopically, we define their observational proxies $\overline{x}^\Delta, \overline{y}^\Delta$ as spatial averages over a box $\Delta$ having linear dimension $\lD$ of the order of the characteristic deep-convection length scale \citep{Pujol2019}. A single deep-convection cell is typically estimated to have a characteristic vertical and horizontal scale $\mathscr{d} \sim \mathscr{l} \sim \SI{10}{\km}$ (fig.~\ref{fig:Lifecycle}). 
We refer to this characteristic length scale as $\lc$. The principal procedure is illustrated by means of example image sequences in figure~\ref{fig:RSexample}. \par
Satellite data are taken from the SEVIRI (Spinning Enhanced Visual and InfraRed Imager) instrument onboard the geostationary Meteosat Second Generation (MSG) satellite \citep{Schmetz2002}. For this study, we use the rapid-scan data of MSG SEVIRI available over Europe at a sampling rate of $\Delta t = \SI{5}{\minute}$ and with a spatial resolution up to $\SI{1}{\km} \times \SI{1}{\km}$ at sub-satellite point. Multi-channel identification algorithms of cumulus convection have been proposed \citep{Berendes2008,Thies2011}. However, as growing cumuli are well discernible already in the high resolution visible (HRV) channel alone, in this study, we consider only the HRV satellite imagery to define $\overline{x}^\Delta$. Our processing routines of the satellite data are based on the open-source library Satpy \citep{Hoese2019}. \par
As in previous studies \citep{Wapler2021,Wilhelm2023}, radar reflectivity data are taken from the RX composite over Germany, operated by the German Meteorological Service \citep{Radolan2024}. Data acquisition is in 256 intensity classes at the same sampling rate of $\Delta t = \SI{5}{\minute}$ and with a spatial resolution of $\SI{1}{\km} \times \SI{1}{\km}$. We use these data to define $\overline{y}^\Delta$. Our radar data processing is based on the open-source library wradlib \citep{Heistermann2013}. \par
In order to have a maximum of available data, we set the observational domain as the polygon with corner coordinates (\ang{7}E, \ang{47}N), (\ang{13}E, \ang{47}N), (\ang{13.5}E, \ang{53.5}N) and (\ang{6.5}E, \ang{53.5}N), roughly corresponding to Germany. All observational data are mapped onto the same $\SI{900}{\km} \times \SI{900}{\km}$ Radolan Cartesian coordinate system $(\hat{x}, \hat{y})$ on the WGS84 earth model. \par
There is of course a multitude of different processes happening in the atmosphere and leaving a trace in the considered remote-sensing observations. In order to single out the relevant data instances, we rely on the definition of thunderstorms to be associated with lightning (Lilly 1979). Lightning observations are available with very high spatiotemporal detection accuracy 
from the ground-based LINET network \citep{Betz2009}. \par
For the purpose of this study, we use the occurrence of a single cloud-to-ground lightning event with an absolute strength of at least \SI{400}{\kilo\ampere} to define that a thunderstorm in the mature stage is currently present in the symmetric box $\Delta$ of linear dimension $\lD$ symmetrically around the central lightning event in $\hat{\bm{x}}_0$ (figure~\ref{fig:RSexample}). This criterion 
was chosen with the emphasis to identify extreme events. As suggested by the heuristic convection lifecycle model (figure~\ref{fig:Lifecycle}), this event is taken to roughly mark the midpoint $t_0$ of the overall evolution. Spatial isotropy and temporal symmetry of the lightning density over the convection lifecycle were shown in previous studies \citep{Wapler2021}. \par
Taking the same box $\Delta$ around the same fixed spatial position for $\pm M$ (here, $M = 48$) data acquisition time steps $\Delta t$ from $t_0$ and spatially averaging the remote-sensing observations over this box yields scalar time series $\overline{x}^\Delta (t), \overline{y}^\Delta(t)$ associated with the model variables $x, y$. Time series are then normalised $\xD_i(t) \leftarrow (\xD_i(t) - \min_t\xD_i)(\max_t\xD_i - \min_t\xD_i)^{-1} \in [0,1]$ ($i = 1,2$), where time $t$ runs from $t_{-M} = t_0 - M\Delta t$ to $t_M = t_0 + M\Delta t$. We call the graph of $t \mapsto (\overline{x}^\Delta (t), \overline{y}^\Delta(t))$ the observational signature associated with the convection lifecycle. \par
Since deep convection essentially denotes the process of thermally driven vertical transport, it is natural to associate a time scale $\tc = \mathscr{u}^{-1}\mathscr{d}$ characteristic of advective processes. With the typical depth and velocity scale $\mathscr{d} = \SI{10}{\km}$ and $\mathscr{u} = \SI{5}{\metre\per\second}$ \citep{Vahid2024}, this yields $\tc = \SI{30}{\minute}$ \citep{Doswell1985}. We emphasise that $\tc$ provides a timescale of the updraught only. An estimate for the entire lifecycle would rather be $2\tc = \SI{60}{\minute}$ \citep{Markowski2010}. \par 
Obviously, extending the time interval to cover $\pm M\Delta t = \pm\SI{240}{\min}$ symmetrically around $t_0$ is much longer than the characteristic convective timescale $\tc$. We have chosen this much longer time interval at this first stage of the analysis deliberately to check our hypothesis of temporal scale separation. For deep-convection events far from equilibrium, any large-scale recovery mechanism (corresponding to the reverse reaction in \eqref{eq:Global_reaction}) must be much slower than the convective timescale. Furthermore, for single-cell events, there should ideally be only one cycle in the observational signal contained in the longer time interval. In either case, the principal signature of a single-cell deep-convection event must be such that the convective variables, starting from small values, increase to much larger values on a timescale of the order of $\tc$ before going to zero again. \par
In order to obtain a statistically representative sample of observational signatures, we draw $N = 10^4$ lightning events at random from the interior of the observational domain from May to October 2018. For the visible satellite imagery to be available for the entire time series, we further restrict the central lightning event $t_0$ to occur between 1100 and 1600 UTC. \par
In our algorithm, the existence of a lightning event is a necessary condition for the occurrence of deep convection. However, it is not sufficient to extract meaningful or useful observational data relative to the phenomenon. According to \citet{Haken1991}, Jaynes' principle states that for the understanding and prediction of a reproducible macroscopic event all microscopic details uncontrolled by the experiment are irrelevant. Hence, our objective in the remainder of this section is to show that the macroscopic deep-convection variables $\xD, \yD$ indeed evolve through a recurrent temporal signature.

\subsection{Discussion of the time-series data set}
\label{sec:Time_series_discussion}
Our model development in section~\ref{sec:Model} assumes a well-mixed system, so that a characterisation in terms of macroscopic variables is justified. We think of deep convection as a coherent flow structure that results from the synergetic interaction of a large number of smaller processes (particles). While each smaller scale process may be very complex, poorly understood and unobservable today (e.g. microphysics), the integrated net effect manifesting as deep convection is observable and can be characterised in terms of macroscopic variables. In order to test our hypothesis of deep convection being a well-mixed coherent flow entity, we systematically varied the box size used for averaging. \par
\begin{figure}
    \centering
    \includegraphics[width=0.9\columnwidth]{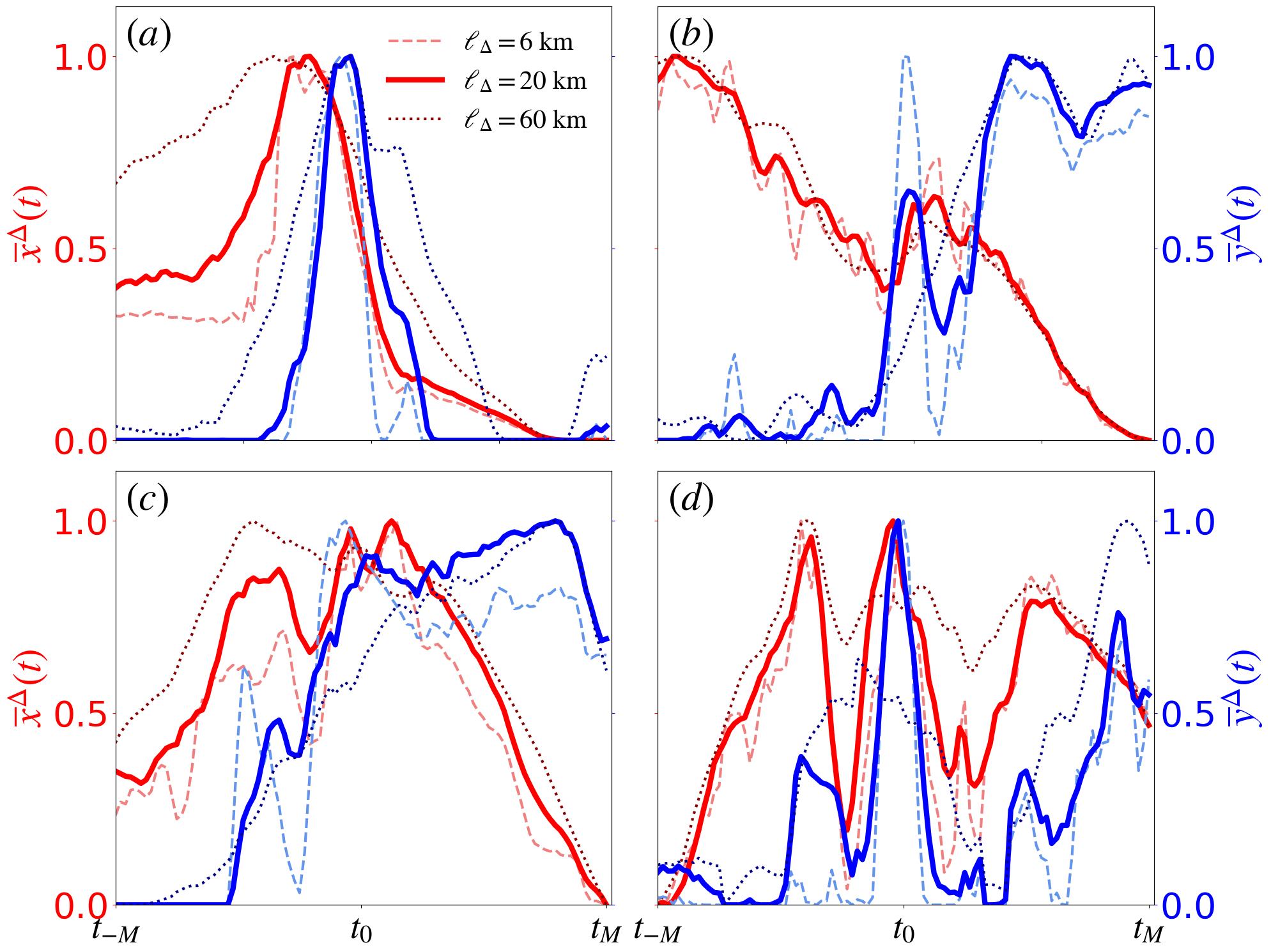}
    \caption{Time-series prototypes for variable averaging box size $\lD\lc^{-1} \in \{0.6,2,6\}$: $(a)$ lifecycle archetype $(b)$ global maximum offset from $t_0$ $(c)$ too broad maximum and $(d)$ noisy. Graphs of $\xD(t)$ and $\yD(t)$ are shown in red and blue.}
    \label{fig:Timeseries}
\end{figure}

 Figure~\ref{fig:Timeseries} shows time series $t \mapsto (\xD(t), \yD(t))$ obtained for linear dimensions $\lD\lc^{-1} \in \{0.6, 2, 6\}$. Indeed, the principle trend apparent is that smaller integration domains lead to pronounced fluctuations of the time series while larger box sizes smooth the signature considerably. In both cases, the disturbing effect is of comparable order to the actual signature itself. This suggests that $\lD\lc^{-1} = 2$ realises a system that is sufficiently extended to average over a large number of constituent elements while being small enough to be characterised in terms of macroscopic variables. The following analyses assume a box size $\lD\lc^{-1} = 2$. \par

As stated above, we take the occurrence of lightning as a necessary condition that deep convection is ongoing. However, this criterion is not sufficient to restrict to single-cell events, nor does it account for possible masking of the signature by processes happening in parallel (e.g. high clouds). Inspection of the random sample reveals that the time series essentially take one out of four qualitatively discernible prototypical signatures shown in figure~\ref{fig:Timeseries}. A rigorous definition and separation of these classes is out of the scope of this study. Rather, we conjecture that there exists a discernible and reproducible temporal signature of deep convection (figure~\ref{fig:Timeseries}$a$), and that all events with this signature form the \textquote{positive} class. The remainder events are then simply grouped into a \textquote{negative} class. \par
Figure~\ref{fig:Timeseries}$a$ corresponds to our expectation of a prototypical convection lifecycle temporal signature: First $\xD$, measuring cumulus cloud deepening, grows rapidly. As the cloud deepens, the precipitation intensity $\yD$ picks up, attaining its maximum temporally lagged compared to $\xD$ \citep{Feingold2013}. The peaks are sharp, and eventually both variables rapidly decay and vanish. The other frequently observed temporal signature are displayed in figure~\ref{fig:Timeseries}$b$--$d$. These correspond to $(b)$ the global maximum being temporally offset significantly from the central time instance $t_0$, $(c)$ very broad, plateauing maxima, and $(d)$ noisy signals with no discernible temporal pattern.

\subsection{Conditional sampling}
\label{sec:Time_series_conditional_sampling}
As our model is designed for a prototypical cell lifecycle, we do not expect it to reproduce individual realisations but only the principal evolution characteristics that hold on average. Also, we do not expect it to reproduce all four generic signatures shown in figure~\ref{fig:Timeseries} but only signature $(a)$, which we assert to correspond to the observational manifestation of the cell lifecycle. Based on the observed qualitative characteristics shown in figure~\ref{fig:Timeseries}, in this section, we introduce a conditional sampling procedure to remove time series from the random sample not covered by our model. \par
Our qualitative discussion of the signatures in figure~\ref{fig:Timeseries} shows that deviations from the positive class (figure~\ref{fig:Timeseries}$a$) are mainly due to two issues. Namely, either the maximum has a considerable lag with respect to the central time $t_0$, or the governing temporal scales differ significantly from the prototypical. The main resulting issue is that the average signature taken over the entire sample is significantly smeared out due to non-isolated or too broad events. We therefore ascertain that the different signatures can be objectively distinguished by recourse to an appropriate conditional sampling.  We therefore implement two post-processing algorithms to guarantee that 
\begin{enumerate*}[label=(\roman*)]
    \item the maximum of the individual time series is close to the average maximum location and
    \item the width of the maximum is not larger than the average.
\end{enumerate*}

For the first criterion, we impose a maximum temporal separation of $5\,\Delta t \sim \tc$ of individual maxima from the average location. That is, if
\begin{equation}\label{eq:First_criterion}
    \arg\max_t \xD_n > 5\Delta t \quad\lor\quad \arg\max_t \yD_n > 5\Delta t
\end{equation}
for the $n$th time series ($n = 1,\ldots,N$), we remove both $\xD_n(t)$ and $\yD_n(t)$ from the random sample. \par
\begin{figure}
    \noindent\includegraphics[width=.9\columnwidth]{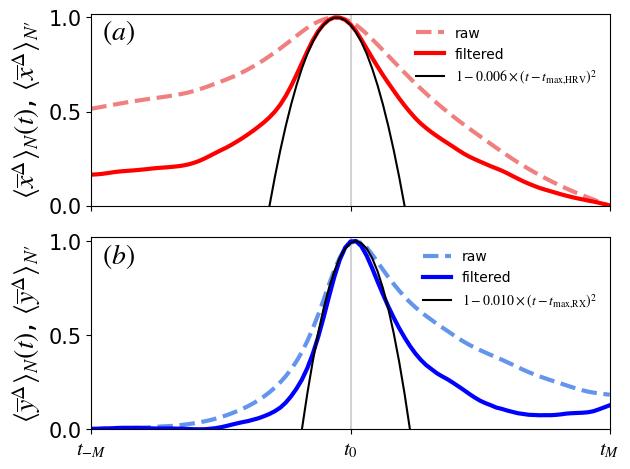}
    \caption{Effect of the conditional-sampling procedure to pronounce the signature of the convection lifecycle in the $(a)$ HRV and $(b)$ RX time series. Time series labelled \textquote{raw} refer to averages taken over the entire random sample of cardinality $N$, $\langle\xD_i\rangle_N(t)$. Those labelled \textquote{filtered} are obtained from averaging over the cleaned sample containing $N'$ elements, $\langle\xD_i\rangle_{N'}(t)$. The parabola model of the convection lifecycle from \citet{Wilhelm2023} is shown in black.}
    \label{fig:Data_cleaning}
\end{figure}

For the second criterion, we use a spectral argument. We reject all time series for which the largest spectral amplitude (denoted $\big|\widehat{\overline{x}^\Delta_i}\big|$) exceeds the mean maximum spectral amplitude computed over the whole random sample. Denoting the ensemble average taken over the entire sample by $\langle\xD_i\rangle_N$ ($i = 1,2$), this criterion becomes: if
\begin{equation}\label{eq:Second_criterion}
    \max_f\big|\widehat{\overline{x}^\Delta_n}\big| > \langle\max_f\big|\widehat{\overline{x}^\Delta}\big|\rangle_N \quad\lor\quad \max_f\big|\widehat{\overline{y}^\Delta_n}\big| > \langle\max_f\big|\widehat{\overline{y}^\Delta}\big|\rangle_N
\end{equation}
for the $n$th time series ($n = 1,\ldots,N$), we remove both $\xD_n(t)$ and $\yD_n(t)$ from the random sample. Applying \eqref{eq:First_criterion} and \eqref{eq:Second_criterion} results in a cleaned sample with $N' = 1344$ elements. The ensemble averages taken over the entire \textquote{raw} and \textquote{filtered} samples are denoted by $\langle\xD_i\rangle_N(t)$ and $\langle\xD_i\rangle_{N'}(t)$, respectively. Figure~\ref{fig:Data_cleaning} displays the effect of the conditional-sampling algorithm on the whole sample. \par
We arrive at the pivotal conclusion that definite recurrent signatures exist in the relevant observational data that can be reproduced in different realisations of the same experiment. By Jaynes' principle, these observational signatures of the deep-convection lifecycle must be explainable and predictable by a macroscopic model. \par

\citet{Wapler2021} conducted a comprehensive statistical analysis of the deep-convection lifecycle characteristics over Germany using radar and lightning data. This study defines the cell lifetime as the elapsed time between the first and last detection of a simply connected region of at least \SI{15}{\km\squared} for which the radar reflectivity attains at least 46~dBZ. In agreement with our above conclusion, \citet{Wapler2021} observed a typical temporal evolution and inferred that the lifecycle of various macroscopic cell attributes should roughly proceed through an inverse parabola, as was already suggested by \citet{Weusthoff2008}. Using our normalisations, the explicit form of this model proposed by \citet{Wilhelm2023} reads
\begin{equation}\label{eq:Lifecycle_parabola}
    \langle\xD_i\rangle_{N'}(t) = 1 - \left(\frac{\mathscr{t}_\mathrm{p}}{2}\right)^{-2}(t-t_{\text{max},\langle\xD_i\rangle_{N'}})^2 \quad (i=1,2) .
\end{equation}
The evolution predicted by \eqref{eq:Lifecycle_parabola} is shown in figure~\ref{fig:Data_cleaning}. We note that the considered macroscopic cell attributes $\xD, \yD$ attain their maximum values at $t_{\text{max},\langle\xD\rangle_{N'}} \approx t_0 - 3\Delta t$ and $t_{\text{max},\langle\yD\rangle_{N'}} \approx t_0$. The width of the parabola \eqref{eq:Lifecycle_parabola} is determined by a characteristic timescale $\mathscr{t}_\mathrm{p}$. The model shown in figure~\ref{fig:Data_cleaning} sets $\mathscr{t}_\mathrm{p}$ to \SI{25}{\min} and \SI{20}{\min} for the satellite imagery and radar data, respectively. These values are close to the characteristic deep-convection timescale $\tc$ and in good agreement with the results of \citet{Wapler2021} and \citet{Wilhelm2023}. \par
Overall, the parabola model \eqref{eq:Lifecycle_parabola} closely reproduces the convection lifecycle in a neighbourhood around the maxima. The temporal range of validity is of the order of the convection timescale \citep[see also][]{Wilhelm2023}. Beyond, agreement drops very rapidly quantitatively and qualitatively. For instance, the lifecylce is not temporally symmetric around $t_\text{max}$ as previously thought \citep{Wapler2021}. Given their definition of the cell lifecycle (recalled above), it is not surprising that the model does not generalise other evolution stages \citep[see also][]{Wilhelm2023}. \par
Eventually, we stress that satellite imagery was not originally considered by \citet{Weusthoff2008}, \citet{Wapler2021} and \citet{Wilhelm2023}. Our results show that the same model may be useful to predict cell attributes derived from satellite imagery. Most notably, our study indicates that the use of satellite imagery has a true added value over radar data in the context of nowcasting deep convection. Figure~\ref{fig:Data_cleaning} shows that deep convection admits a recognisable and recurrent signature in satellite imagery preceding the corresponding radar signature by around $\SI{15}{\min}$. This suggests that the predictability of deep convection can be increased accordingly upon using satellite imagery. \par




\section{Comparison between model and observations} 
\label{sec:Model_observation_comparison}
 
\begin{figure}
    \centering
    \noindent\includegraphics[width=.9\columnwidth]{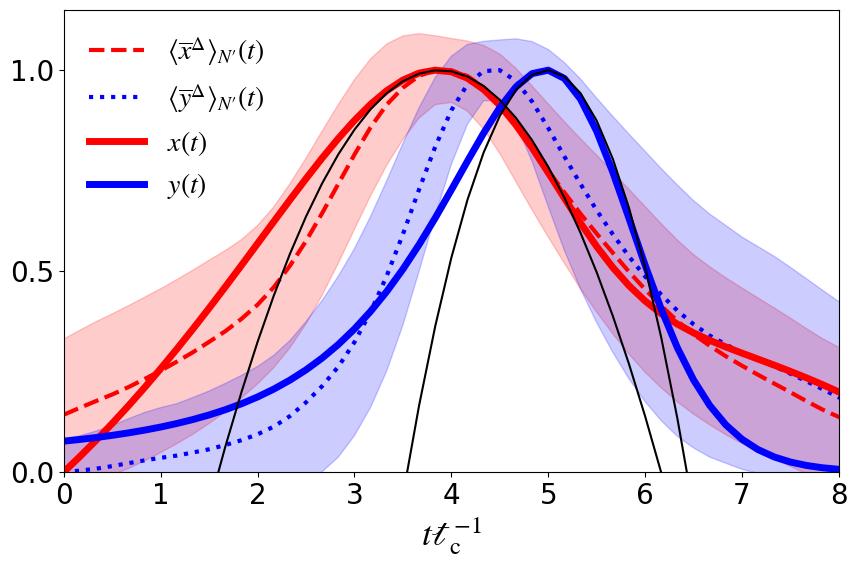}
    \caption{Comparison between the temporal evolution of the convection-lifecycle model and the remote-sensing signature after data cleaning. Model parameters are $\alpha = 0.29$, $\beta = 0.30$, $\gamma = 0.08$, $\delta = 0.52$, $\epsilon = 0.01$ and the integration was started from $\bm{x}_0 = (1.0,0.1,0.1)$. Shading indicates the symmetric range of one standard deviation computed from the filtered ensemble comprising $N'$ elements. The parabola model \eqref{eq:Lifecycle_parabola} is shown in black.}
    \label{fig:Model_signal}
\end{figure}

We have developed our model on the basis of empirical prior knowledge, which suggests a minimum of qualitative consistency with actual manifestations of deep convection in the atmosphere (section~\ref{sec:Model}). In section~\ref{sec:Data}, we proposed a conditional-sampling algorithm of remote-sensing observations to obtain definite recurrent signatures associated with single-cell deep convection. In this section, we demonstrate that our model is capable of reproducing the principal features of the entire convection lifecycle as it manifests in observational data. Compared to the parabola model, this includes the non-symmetric tails far from $t_0$. The comparison is shown in figure~\ref{fig:Model_signal}. \par

Inspection of the time series in figure~\ref{fig:Timeseries}$a$ and \ref{fig:Data_cleaning} shows that the time interval $[t_{-M},t_M]$ was chosen too long for the actual lifecycle (see discussion in section~\ref{sec:Data}\ref{sec:Time_series_construction}). For the quantitative comparison with our model in this section, we therefore restrict to the shorter time interval ranging from $t_0 - 26\Delta t$ to $t_0 + 32\Delta t$. Since we discuss the initial-value problem \eqref{eq:Formal_ode}, we reformat the time to start at zero and non-dimensionalise with the characteristic time scale $\tc$. \par
In order to obtain the model parameters, we use the Nelder--Maed optimisation method \citep{Nelder1965} assuming an $\mathscr{l}^2$-norm for the objective function. Initialising the optimisation with a good hand fit ($\alpha_0 = 0.40$, $\beta_ 0 = 0.10$, $\gamma_0 = 0.25$, $\delta_0 = 0.50$) yields $\alpha = 0.29$, $\beta = 0.30$, $\gamma = 0.08$, $\delta = 0.52$. We find this parameter set to be quite robust with respect to modifications of the initial guess. This relative insensitivity to parameter variations suggests that our model is structurally stable, and hence, robust. Analogously to $\xD, \yD$ (section~\ref{sec:Data}), the model trajectories are normalised $x_i(t) \leftarrow (x_i(t) - \min_t x_i)(\max_t x_i - \min_t x_i)^{-1}$ ($i = 1,2$). Figure~\ref{fig:Model_signal} shows that the dynamics eventually tend to fixed point $\bm{x}_3^*$ (inactive state) as conjectured in section~\ref{sec:Model}\ref{sec:Model_discussion}.  \par
Since $x(t) = O(1)$ for all $t$ and $\alpha \approx \beta$, cumulus growth and the conversion to precipitation happen on similar timescales. On the other hand, $x(t) \sim y(t)$ for all $t$, while $\gamma \ll \delta$. This suggests that cumulus convection $x$ is less efficient at dissipating active atmospheric state than precipitation $y$. \par
Figure~\ref{fig:Model_signal} compares the model trajectories $x(t), y(t)$ with the observational signature averaged over the filtered ensemble (section~\ref{sec:Data}). An estimation of the variability in the observational data is given by the symmetric one-standard-deviation range. Apart from a small deviation around $t\tc^{-1} \approx 4$, our model consistently stays within the range of one standard deviation from the mean signature. We notice that our model correctly reproduces the temporally lagged cumulus cloud growth followed by precipitation as emphasised in several previous studies \citep[e.g.][]{Davies2009,Koren2011,Koren2017,Pujol2019}. Also, superposed in figure~\ref{fig:Model_signal} is the parabola model \eqref{eq:Lifecycle_parabola}. Around the maxima, the two models practically coincide. However, our model provides a better agreement of the initiation and dissipation stages not covered by \eqref{eq:Lifecycle_parabola}. These regimes are actually associated with the greatest nowcasting errors \citep{Wang2017}. \par
Generally, very close agreement could be obtained by adding additional reactions or species to our model. However, this is out of the scope of this study and principally prone to over-fitting.



\section{Conclusion} 
\label{sec:Conclusion}

Deep convection is a canonical process of atmospheric dynamics and associated with hazardous weather. As such, its understanding and prediction is of major importance for weather and climate modeling as well as forecasting on all scales. The elementary building block of deep convection is a recurrent coherent flow structure referred to as cell, which evolves through its characteristic lifecycle. \par
In this study, we developed a minimal model of the convection lifecycle by formalising the standard conceptual model in the form of reaction rules between four macroscopic variables. Assuming a well-mixed system, this set of reactions is equivalent to a system of nonlinearly coupled differential equations that model the convection lifecycle. \par
We then introduced a conditional sampling strategy to construct and select time series from remote-sensing observations (visible satellite imagery and radar reflectivity) that are associated with manifestations of deep convection in the atmosphere. In particular, we showed that a recurrent signature of the temporal evolution of the convection lifecycle exists. By Jayne's principle, reproducibility of macroscopic phenomena implies that their evolution can principally be predicted by means of the resolved macroscopic variables. \par
Eventually, we demonstrated that our model is capable of reproducing the essential features of the observational signatures qualitatively and quantitatively. In particular, fixing the model parameters by a Nelder--Maed optimisation, the model trajectory essentially stays within one standard deviation from the observational mean. \par
The present work is intended to be a proof of concept to show that the adopted modelling framework is principally useful and that already a simple minimal model may encode the governing dynamics. Conceptually, extensions of the presented model can straightforwardly be developed by including additional or modifying the proposed reactions. By this, spatiotemporal convection organisation (e.g. multi cells, squall lines) or coupling to other atmospheric processes (e.g. fronts, convergence lines) may be modelled. \par
In this work, we have simply and directly identified the (abstract) model variables with remote-sensing observations. Future work should explore multi-channel data to define observable proxies of the model variables. Besides expecting a closer correspondence between $x_i$ and $\xD_i$ ($i=1,2$), we expect this to result in an increased reproducibility of the lifecycle signature. \par 
While we hope that the model will prove useful in general, the application in nowcasting can readily be anticipated. In particular, the Nelder--Maed optimisation can be modified to set the free model parameters from past observations on the fly. The quantitative performance can likely be increased by defining $\xD_i$ from multi-channel observations. This can be combined in an integral physics-informed machine-learning model building. Recently, the usefulness of closely related population-dynamics approaches to develop convection parametrisations has been shown. This readily suggests that our model, or extensions thereof that can straightforwardly and systematically be developed in the framework of reaction schemes, may be adapted for convection parametrisations.

\acknowledgments
This work was done as part of the Italia--Deutschland Science-4-Services (IDEA-S4S) network. C.M. acknowledges funding from this network. We would like to thank R.~Müller from DWD as well as G.~Kilroy, A.~Schäfler and J.~Thayer from DLR for several fruitful discussions and suggestions on earlier versions of the manuscript. \par
The authors declare no conflict of interest.

%
%
\datastatement
Data availability is currently checked with the data providers. A final statement will be added later.

%







%
%
%
\bibliographystyle{ametsoc2014}
\bibliography{main.bbl}

%

%

\end{document}